  \documentclass[10pt,prb,superscriptaddress,showpacs,amsfonts,amsmath,amssymb,showkeys,floatfix,notitlepage,nobibnotes]{revtex4-1}
\usepackage{amsfonts,amsmath,amssymb,graphicx, placeins, bm}
\usepackage{setspace,color,cases}
\usepackage{pstricks}
\usepackage{fdsymbol}

\usepackage{ae,aeguill}
\usepackage[utf8]{inputenc}    
\usepackage[T1]{fontenc}
\newcommand{\la}{\lambda}    
  \newcommand{\ep}{\epsilon} \newcommand{\s}{\sigma}
     
 \newcommand{\tend}{\rightarrow}
\newcommand{\equa}[1]{\begin{eqnarray} \label{#1}}
\newcommand{\auqe}{\end{eqnarray}}
\newcommand{\equab}[1]{\begin{widetext}\begin{eqnarray} \label{#1}}
\newcommand{\auqeb}{\end{eqnarray}\end{widetext}}
\newcommand{\tab}[1]{\begin{tabular}{#1}}
\newcommand{\bat}{\end{tabular} \\ }
\begin {document}
\title 
 {On the phase diagram of a three-dimensional dipolar model.}

\author{V. Russier}
\email[e-mail address: ] {russier@icmpe.cnrs.fr}
\affiliation{ICMPE, UMR 7182 CNRS and UPE 2-8 rue Henri Dunant 94320 Thiais, France.}
\author{Juan J. Alonso}
\email[e-mail address: ] {jjalonso@uma.es}
\affiliation{F\'{\i}sica Aplicada I, Universidad de M\'alaga, 29071 M\'alaga, Spain}
\affiliation{Instituto Carlos I de F\'{\i}sica Te\'orica y Computacional,  Universidad de M\'alaga, 29071 M\' alaga, Spain}

\date{\today}   
\begin {abstract}
 The magnetic phase diagram at zero external field of an ensemble of dipoles with uniaxial
anisotropy on a FCC lattice has been investigated from tempered Monte Carlo simulations.
The uniaxial anisotropy is characterized by a random distribution of easy axes and its
magnitude $\la_u$ is the driving force of disorder and consequently frustration. 
The phase diagram, separating the paramagnetic, ferromagnetic and spin-glass regions, 
was thus considered in the temperature, $\la_u$ plane. 
Here we interpret this phase diagram in terms of the more convenient variables namely the bare 
dipolar interaction and anisotropy energies $\ep_d$ and $\ep_u$ on the one hand and 
the volume fraction $\Phi$ on the other hand and compare the result with that corresponding 
to the random distribution of particles in the absence of anisotropy. 
We also display the nature of the ordered phase reached at low temperature by the ensemble of 
dipoles on the FCC lattice in terms of both the dipolar coupling and the texturation of the 
easy axes distribution when the latter is no more random.
This system is aimed at modeling the magnetic phase diagram of supracrystals of 
magnetic nanoparticles. 
\end{abstract}
\maketitle
\begin {figure}[h!]
  \includegraphics [width = 0.75\textwidth , angle =  00.00]{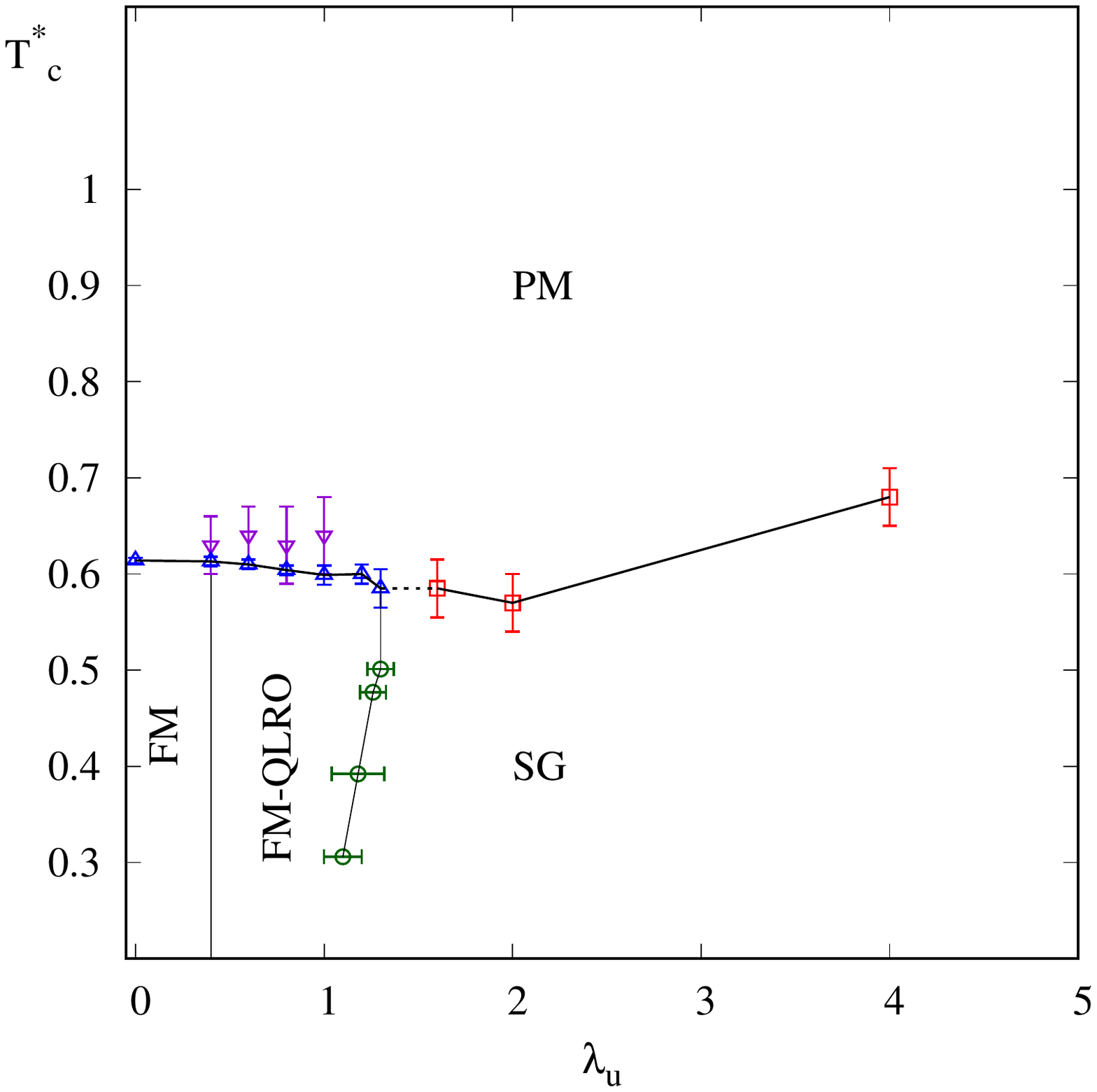}
  \vskip -0.04\textheight
  \caption {\label {ph_diag_lau}
  ($T_c/T_r,\la_u$) phase diagram of the dipolar plus uniaxial anisotropy with random distribution of easy axes
  on a FCC lattice. $\la_u$ is the natural disorder control parameter.
     }
  \end {figure}
\begin {figure}[h!]
  \includegraphics [width = 0.75\textwidth , angle =  00.00]{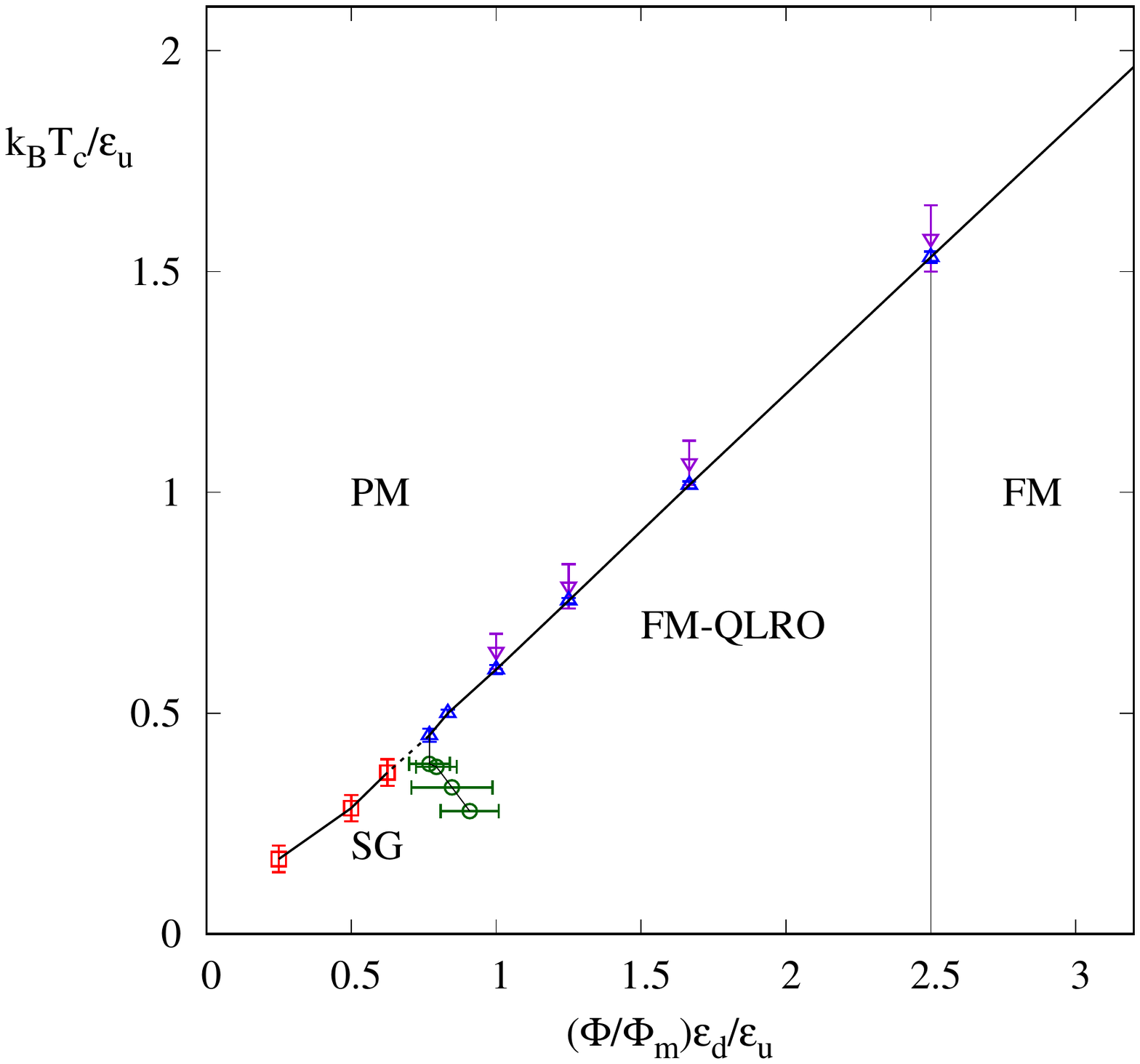}
  \vskip -0.04\textheight
  \caption {\label {ph_diag_phi1}
  Phase diagram of figure~(\ref{ph_diag_lau}), shown as $T_c$ in terms of increasing dipolar coupling
  for $\ep_d$ and $\ep_u$ considered as constant. 
     }
  \end {figure}
\begin {figure}[h!]
  \includegraphics [width = 1.00\textwidth , angle =  00.00]{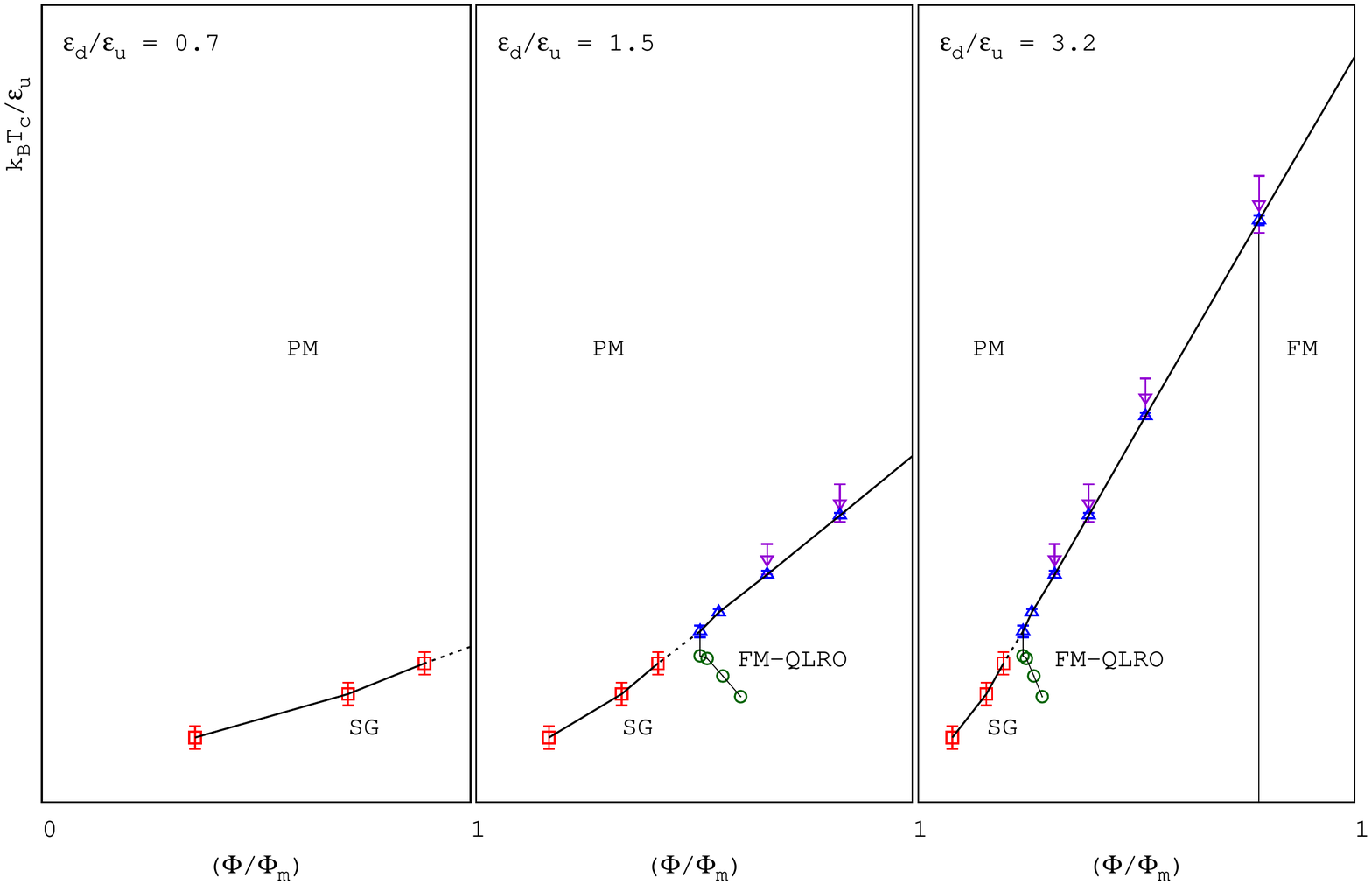}
  \vskip -0.04\textheight
  \caption {\label {ph_diag_phi2}
  Same as figure~(\ref{ph_diag_phi1}) where the constraint $\Phi/\Phi_m\leq{1}$ is explicited for three 
  characterisitc values of $\ep_d/\ep_u$ as indicated. 
     }
  \end {figure}
\begin {figure}[h!]
  \includegraphics [width = 0.60\textwidth , angle = 00.00]{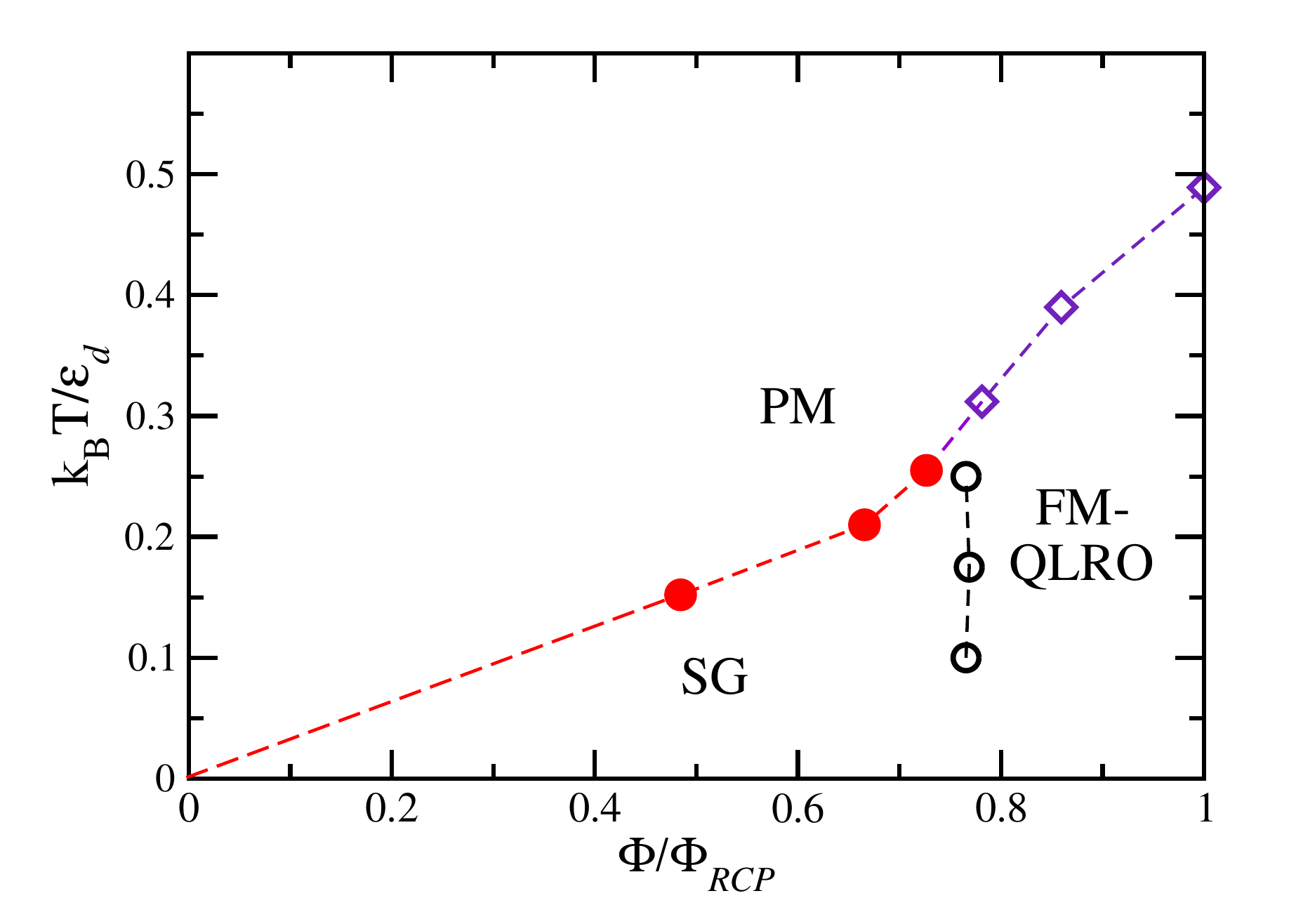}
  \caption {\label {ph_diag_rcp}
  Phase diagram of the dipolar hard spheres distributed according to a frozen hard-sphere like 
  distribution with $\ep_u=0$. The limiting value of the volumefraction, $\Phi_{RCP}$ is indicated.
     }
  \end {figure}
\begin {figure}[h!]
  \includegraphics [width = 0.80\textwidth , angle =  00.00]{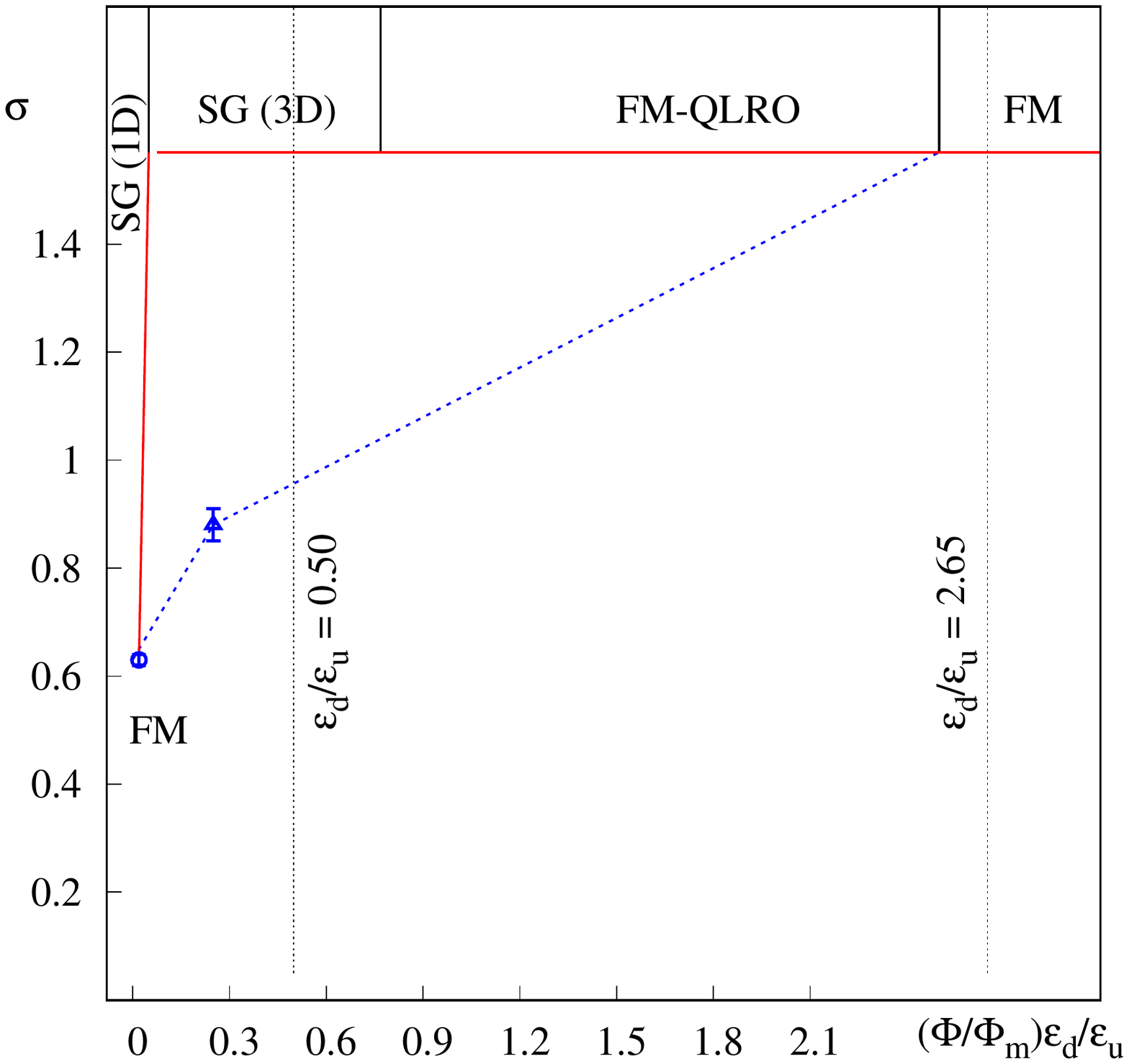}
  \vskip -0.04\textheight
  \caption {\label {ordered_ph}
  Nature of the ordered phase at low temperature for the FCC lattice in terms of the texturation
  quantified by the variance $\s$ of the easy axes distribution (see Ref.~\cite{alonso_2019,russier_2020a})
  and the dipolar coupling $(\Phi/\Phi_m)\ep_d/\ep_u$. Open circle and open triangle are the 
  FM/SG transition in the dipolar Ising limit~\cite{russier_2020a}, and 
  $(\Phi/\Phi_m)\ep_d/\ep_u= 0.25$~\cite{russier_2020c} respectively. The dotted line is only indicative. 
  The upper horizontal red line, $\s=\pi/2$ indicates the lower bound of random distribution of easy axes.
  The SG region is separated in the true dipolar SG (3D) and the limiting dipolar Ising SG (1D) obtained
  for $\la_u>30$~\cite{russier_2020b}.
  The reachable region is located on the left hand side of the line $(\Phi/\Phi_m)\ep_d/\ep_{u}=\ep_d/\ep_u$
  which is shown for a weak and a strong dipolar couplings  as particular cases.
     }
  \end {figure}

%
  In Ref.\cite{russier_2020b} we investigated the phase diagram of an ensemble of dipolar 
hard spheres located on the nodes of a perfect FCC lattice. The phase diagram 
was investigated in the reduced temperature, disorder control parameter, 
($T^*,\la_u$) plane. Given the $1/r^3$ dependence of the dipole-dipole 
interaction (DDI), it has been found convenient to define the reduced temperature, 
$T^*=T/T_r$ through the reference temperature $T_r=(\Phi/\Phi_m)(\ep_d/k_B)$ where
$\Phi$, $\Phi_m$ and $\ep_d$ are the volume fraction, its maximum value of the 
FCC lattice and the dipole-dipole energy for particles at contact.
The disorder control parameter is nothing but the anisotropy energy (MAE) over DDI 
ratio, namely $\la_u=(\ep_u/\ep_d)(\Phi_m/\Phi)$. Then the ($T^*,\la_u$) phase diagram, 
displayed on figure~(\ref{ph_diag_lau}), is qualitatively similar to those corresponding 
to the $\pm{}J$ Ising model~\cite{hasenbusch_2007b,papakonstantinou_2013}. 
The essential feature is the succession at low temperature of long range order FM, quasi 
long range FM (QLRO-FM) plus transverse SG and SG phases with increasing values of $\la_u$. 
Now, when considering the system as a model for the magnetic phase diagram of
supracrystals of 
bare magnetic nanoparticles characterized by given values of $\ep_d$ and $\ep_u$ 
where the volume fraction $\phi$ can be tuned through the thickness of the coating layer,
it is convenient to translate the phase diagram in terms 
of $(k_{B}T/\ep_u)$ and $(\Phi/\Phi_m)(\ep_d/\ep_u)$, the latter being understood 
as the increasing dipolar coupling through the increase of concentration at constant
$\ep_d$ and $\ep_u$. This is shown on figure~(\ref{ph_diag_phi1}).
In this represention, the dipolar lattice phase diagram appears qualitatively similar
to the one deduced from experiments on the DMIM samples~\cite{petracic_2010}.
Furthermore one must limit the reachable part of this phase diagram to 
the region defined by $\Phi/\Phi_m\leq{}1$. Doing this we see, as diplayed on
figure~(\ref{ph_diag_phi2}), that according to the value of ($\ep_d/\ep_u$), 
namely the relative importance of the dipolar coupling to the anisotropy energy, 
the FM-QLRO an the FM phases may be not reachable. In the limit $\ep_d/\ep_u\tend\infty$, 
one gets only the PM/FM transition of the pure dipolar system on the FCC lattice free of 
anisotropy ($\ep_u=0$)~\cite{bouchaud_1993,russier_2017,russier_2020b} and all the 
non trivial features of the phase diagram are shrinked on the $\Phi=0$ line.
Moreover, the ($(k_{B}T/\ep_u),(\Phi/\Phi_m)(\ep_d/\ep_u)$) phase diagram 
for particles located on the perfect FCC lattice can be directly compared to 
the one we have got~\cite{alonso_2020a} for the system of dipolar spheres frozen in 
a random hard sphere like distribution, in the absence of anisotropy, where the increasing
disorder is quantified by $1-(\phi/\phi_{RPC})$, the volume fraction being then limited to 
the so-called RCP value, $\Phi_{RCP}\simeq{}0.64$, given on figure~(\ref{ph_diag_rcp}). 
It is also qualitively close to the one of dipolar Ising model on the SC lattice in terms
of dilution~\cite{alonso_2010}, where however the ordered phase is anti-FM due to the
lattice symmetry.
Finally, we present on figure~(\ref{ordered_ph}) the nature of the ordered phase
at low temperature in terms of the dipolar coupling and the texturation of the 
easy axes distribution as was studied in~\cite{alonso_2019,russier_2020a} in the 
limit of infinitely strong anisotropy energy leading to the dipolar Ising model. 
As is the case above, the condition $\Phi/\Phi_{m}\leq{1}$ restricts the diagram 
to the left hand part defined by the abcissa $\leq(\ep_d/\ep_u)$ and has been displayed
for two particular cases corresponding to a weak and a strong dipolar coupling respectively. 
  %
  \vskip -0.04\textheight
  %
  %
  \vskip -0.04\textheight
  %
  %
%
%
\end   {document}